\documentclass{jjap3}

\title{
A Theoretical Simulation of Deformed 
Carbon Nanotubes with Adsorbed Metal Atoms: 
Enhanced Reactivity by Deformation 
}

\author{Naoki Hosoya, 
Koichi Kusakabe\thanks{E-mail address: kabe@mp.es.osaka-u.jp}, and 
S. Uma Maheswari$^{1}$}

\inst{Graduate School of Engineering Science, Osaka University, 
1-3 Machikaneyama-cho, Toyonaka, Osaka 560-8531, Japan \\
$^{1}$Meenakshi College for Women, 
Acrot Road, Kodambakkam, Chennai 600 024, India
}

\abst{
First-principles simulations were performed to investigate 
reaction of carbon nanotubes with adsorbed metal atoms. 
Mechanical modification of their structures enhances 
chemical reactivity of carbon nanotubes. 
Adsorption of a tungsten, tantalum, or niobium atom
on a (5, 0) nanotube with a Stone-Wales defect was 
shown to have characteristically strong chemisorption.
Bond-breaking in the carbon-carbon network and formation
of a local metal-carbon complex were observed during the simulation.
Adsorption of W, Ta, Ni or Mo on a twisted (5, 0) nanotube 
showed a preferred breaking 
of several bonds, even creating an opening in the wall.
The enhanced chemical reactivity of deformed nanotubes
is characterized by formation of a metal-carbon complex.
Applications of the reaction are suggested. 
}

\begin{document}
\maketitle

\section{Introduction}

To investigate chemical nature of the carbon nanotube\cite{Iijima} 
against reactivity to metal species is important 
to understand physical properties of the nanotubes. 
When this material is placed in an environment 
with other metallic elements, 
carbon atoms can react with metal atoms. 
This environment is found in various situations. 
In the catalyzed synthesis of nanotubes, formation of 
metal-carbon composites is known to be essential.\cite{Thess} 
Once the nanotube grows up, however, 
the tube tends to keep its independent 
structure from the metal catalysts. 
People often create metal-nanotube interfaces or metal-nanotube 
junctions,\cite{Tans,Martel} 
when the electronic devices are fabricated using the nanotube. 
Chemical reactivity of metal on the wall 
of the carbon nanotube at the interface 
has been studied by using both experimental\cite{Zhang,Jin-Phillipp,Yang2} 
and theoretical approaches.\cite{Durgun,Yang1,Zhuang,Duong,Inoue,David} 
Selection of metal elements is a key to control the reaction. 
This big issue has been under discussion. 

When a theoretical simulations is applied, 
generally speaking, the reactivity of a metal atom on 
the pristine carbon nanotube with perfect wall does 
not always look so reactive.\cite{Durgun} 
Compared to this knowledge, 
real experiments often show 
an apparent high reactivity.\cite{Zhang,Jin-Phillipp,Yang2} 
One reason is the temperature effect,\cite{Duong,Inoue} 
by which the tube walls are disturbed and become reactive. 
Another is local deformation from a perfect wall. 
Insertion of defects and creation of strained structures are 
typical examples.\cite{Yang1,Zhuang,David} 
To investigate origin of enhancement in the chemical reactivity, 
a microscopic simulation using some deformed nanotubes can be a good tool. 
This idea is actually utilized in the above-mentioned researches. 
If we further add information on a chemical trend on elements, 
we would be able to open a new scope. 

In this study, we focused the chemical reactivity of 
some specified nanotubes with metal atoms. 
For this purpose, we selected a pristine (5,0) tube 
as a starting point for our discussion. 
Comparison among some late transition elements tells that 
tungsten and tantalum could show reactivity qualitatively 
different from others. 
To make apparent clues, we considered deformed nanotubes. 
One is a tube with the Stone-Wales defect. 
The other is a twisted tube, which is prepared in a computer simulation. 
A big difference between some late transition elements and 
the noble metal elements is found, when we consider 
reactivity on the deformed nanotubes. 
We will focus on formation of metal-carbon complex 
to understand this chemical trend. 

\section{Calculation conditions}

To find high chemical reactivity in some mechanically deformed nanotubes, 
we performed the electronic structure calculation of 
the nanotube using methods in the density functional 
theory.\cite{Hohenberg,Kohn} 
We searched for enhanced reactivity in simulations 
to optimize the atomic structure of a metal adsorbed nanotube. 
To have this optimized structure, we applied the Born-Oppenheimer 
dynamics realized in the Quantum-Espresso package.\cite{Giannozzi}
A generalized-gradient approximation (GGA) is 
applied\cite{Perdew2,Perdew3} 
for the description of the exchange-correlation energy functional, 
and each ionic potential for the valence electrons is 
given by the ultra-soft pseudo-potential.\cite{Vanderbilt}

Major parameters in the simulation are summarized here. 
The energy cut-off $E_{c}$ in the plane-wave expansion 
for the wave function is 20[Ry] or higher. 
The $k$-mesh size is 1$\times$1$\times$10 for all simulation cells 
with the $z$ direction parallel to the tube axis. 
A criterion to check convergence is 
to stop simulation when the total absolute value of 
the inter-atomic force becomes less than 
$F_{c}=1.0 \times 10^{-4}$[Ry/a.u.]. 
The length of the nanotube is optimized by finding 
an energy minimum with respect to 
the simulation cell size along the tube axis. 

\section{Preparation of defects and deformation of the nanotube}

The defects of the carbon nanotube may be classified into three categories: 
i) topological defect, ii) re-hybridization defects, and iii) 
incomplete bonding.\cite{Ebbessen,Charlier-rev,Jiang} 
Among these defects, the Stone-Wales defect (SWD)\cite{Stone} 
attracts much attention. 
Motion of carbon atoms 
to form SWD has been observed,\cite{Suenaga} 
so that the defect exists in real nanotubes. 
Strain release and bending deformation are often explained 
as results of formation of 
SWD,\cite{Nardelli1,Nardelli2,Terrones,Chandra,Wako,Pozrikidis}
because SWD may create an essential dislocation center of 
the tube wall. 
Thus the topological defect is relevant when we consider 
mechanical deformation of the tubes. 

These defects or defective sites modify the electronic structure of 
the carbon nanotube.\cite{Amelinckx,Charlier,Carroll,Kim,Meunier,Tekleab} 
As shown in these previous studies, 
the electronic states around the Fermi level is modified by defects. 
Thus, existence of defects may change chemical reactivity of 
the nanotube.\cite{Yang-Shin-Kang,Horner} 

Mechanical modification, which can effect defect formation,
can also influence chemical properties of the carbon nanotubes. 
Mechanical deformation is especially interesting owing to the 
morphological flexibility of the nanotube 
against external stress.\cite{Yakobson,Wong} 
Many characteristic mechanical properties 
are known to originate from special defects 
formed by non-6-membered rings in the nanotube. 

Thus we consider a nanotube with SWD and a twisted nanotube. 
Further we focus on thin tubes, {\it e.g.} the (5,0) nanotube, 
to have enhanced reactivity. 
Enhancement is expected for thinner tubes than thicker tubes 
as discussed by Seo {\it et al.} on NO$_2$ adsorption.\cite{Seo} 
The (5,0) tube is known to be metallic.\cite{Miyake} 
On the (5,0) nanotube wall, we attach metal atoms. 
Chemical reactivity is considered by finding chemical trend 
and by showing a method enhancing the reactivity. 

We prepare these structures of carbon nanotubes. 
An optimized structure with 40 carbon atoms is prepared in 
a super cell of 15 $\times$ 15 $\times$ 8.5 \AA. 
To obtain the tube with defects, we use another cell 
whose size is  12.3 $\times$ 12.3 $\times$ 12.8 \AA, and 
which contains 60 carbon atoms. 
The value of 12.8 \AA \, is obtained by finding the energy minimum 
of the pristine (5,0) tube. 
In this cell, a tube with a Stone-Wales transformation is prepared 
as well as the pristine (5,0) tube. 
The optimum value of the cell along the tube axis 
becomes 12.9 \AA, which indicates that 
the tube with SWD becomes slightly longer than the pristine tube. 
The total energy of this nanotube with SWD is 
higher than the pristine (5,0) tube by 0.056[eV] 
per carbon atom. This value is consistent with known creation energy of 
single SWD, which is around 3$\sim$ 5.5[eV].\cite{Conversano,Zhou,Picozzi,Li}

A twisted (5,0) tube is also prepared using this super cell with 
60 carbon atoms. 
To determine the twisted structure, an initial configuration 
of carbon atoms is given by uniformly rotating atomic positions 
around the tube axis. The angle of the rotation is increased 
along the $z$ direction. 
By applying the structural optimization, 
a twisted nanotube is obtained. 
The tube axis again becomes longer and the optimum value of 
the cell along the tube axis becomes 13.0 \AA. 
The structure obtained has a little flattened section, 
whose deformation pattern is similar to 
those found in a former study.\cite{Pozrikidis} 
The obtained structure has a total energy of 
0.43[eV] per carbon atom higher than the pristine (5,0) tube. 

\section{Single atom adsorption on a pristine (5,0) nanotube}

We considered adsorption sites for various atoms on 
a pristine (5,0) tube. 
A super cell with 40 carbon atoms was used. 
These simulations were performed to check the 
accuracy of the calculation at first, 
but we may be able to find 
difference in results obtained for other tubes. 
In this work, 25 elements including 
Ca, Sc, Ti, V, Cr, Mn, Fe, Co, Ni, Cu, Zn, 
Y, Zr, Nb, Mo, Ru, Rh, Ag, Cd, 
Ta, W, Re, Os, Pt, and Au, were tested. 

The adsorption sites were almost the same as those found 
for the (8,0) tube.\cite{Durgun} 
Here, although we performed only spin-unpolarized calculations for each 
structural optimization for the first test, we concluded that 
our simulation results preserved reasonable accuracy. 
Difference in adsorption sites were found for Co, Cu, Ag, and Zn. 
Stable sites were the H site for Co, the A site for Cu and Ag, while 
we could not find any stable site for Zn on the (5,0) tube. 
Here the H site is a hollow site just above the center of a hexagon, 
and the A site is a position above an axial carbon-carbon 
(CC) bond. 
In the previous study by Durgun {\it et al.}, 
Co and Zn preferred the H site, and Cu and Ag went to the A site 
on the (8,0) tube. 
The difference in the stable adsorption site is interesting in itself, 
but we searched for a more apparent qualitative difference. 

\begin{figure}
\begin{center}
(a) 
\includegraphics[width=0.5\hsize]{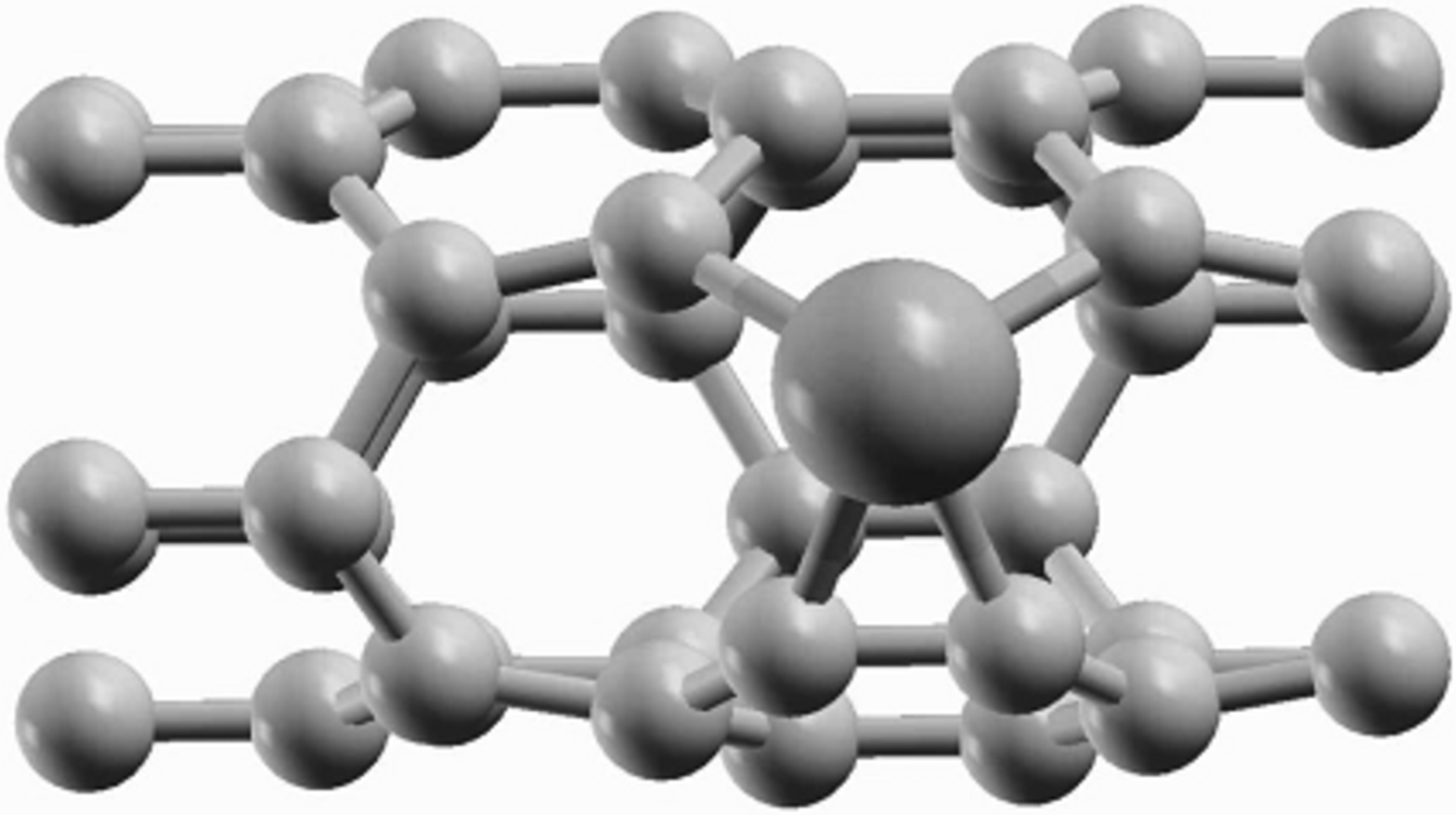}\\
(b) 
\includegraphics[width=0.5\hsize]{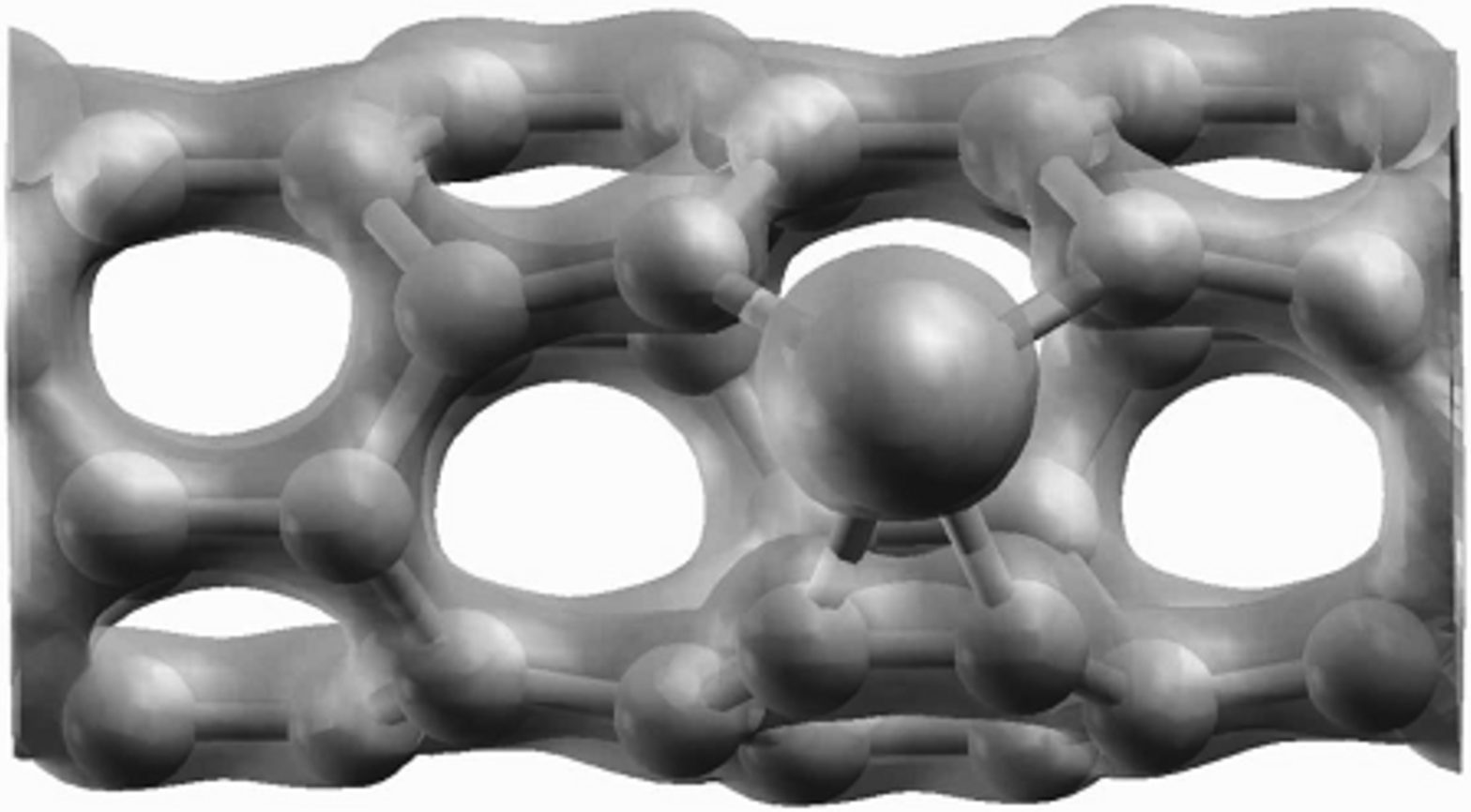}
\end{center}
\caption{(a) Adsorption of a tungsten atom on the (5.0) nanotube. 
When spin-unpolarized GGA is utilized, the adsorbed W atom (a big sphere) 
breaks two carbon-carbon bonds and forms a strong chemisorbed structure. 
In (b), iso-surface of the charge distribution is overwritten. }
\label{f1}
\end{figure}

For selected 10 elements, by increasing $E_c$ to 30 Ry, 
we performed convergence check, which 
results in a little shifted atomic positions 
for most of the calculations. 
The convergence criterion for the inter-atomic force 
was $1.0\times 10^{-4}$ [Ry/a.u.]. 
But, when a tungsten adsorbed (5,0) tube was 
optimized carefully, the obtained atomic configuration 
showed a chemisorbed structure with remarkable characteristics. 
In Fig.~\ref{f1}, the obtained (5,0) tube with W is shown. 
The atomic position of W shifts from the center of hexagon, 
four carbon atoms come close to the W atom, 
several chemical bondings in the tube wall are greatly modified, 
and two CC bonds are significantly weakened. 
Similar rebonding in the metal adsorbed tube 
was seen also in a tantalum-adsorbed (5,0) tube. 
The elongated bond length of the CC bond 
just around the adsorbed metal was 
2.57 \AA, and 1.63 \AA, for tungsten and tantalum, respectively. 
Clear shift in the adsorbed metal atom from the symmetric H site 
to a site surrounded by four C atoms 
was also seen for Mo and Fe. 

We may assign the curious chemisorption as a specific feature 
of the (5,0) tube. The calculation condition of the spin unpolarized 
approximation might cause difference from a spin-polarized solution. 
Actually, the spin-polarized GGA recalculation of the (5,0) tube 
with a W atom did not show similar strong bond cuttings, 
although a shifted adsorption site from the H site was seen. 
In a case of W adsorption on a (5,5) tube, 
the W atom went to the H site, and 
we had no clear sign of the bond breaking. 
However, the above strong chemical rebonding found in the simulation 
of the thin (5,0) tube with W can be a first sign of 
possible strong chemical reactivity of metal atoms 
on some active nanotube walls. 
Thus, we next considered a way to enhance reactivity of the tube wall 
by insertion of defects and by introduction of mechanical modifications. 

\section{Adsorption on a tube with a Stone-Wales defect}

We can see another evidence that some transition metal elements react with 
the tube wall rather strongly. 
Adsorption of a tungsten atom was next investigated using 
the nanotube with SWD. 
The W atom was initially placed 
3 \AA above the tube around a 7-membered ring. 
The optimized structure again showed formation of 
a tungsten-carbon complex. 
Inter-atomic distances are summarized in Table~\ref{t1}. 
The CC bonds become rather long around the adsorbed site. 
The same tendency is found for Nb and Ta, too. 
The mean distance from an adsorbed metal atom to neighboring carbon atoms 
represented as CM is also evaluated. (See Table~\ref{t1}.) 
The values are even shorter than CC1 and CC2, suggesting 
rehybridization of atoms. 
The charge density profile for the optimized structures tells that 
covalent charge almost disappears at elongated CC bondings. 
For W, Nb, and Ta, a four-fold coordination 
of the metal atom seems to be favored. 
These characteristics for the formation of a complex is not 
seen, when the adsorbed metal is platinum. 
Thus we conclude that, when a selected transition metal is adsorbed 
a metal-carbon complex is created. 

\section{Adsorption on a twisted nanotube}

\begin{figure}
\begin{center}
(a) 
\includegraphics[width=0.8\hsize]{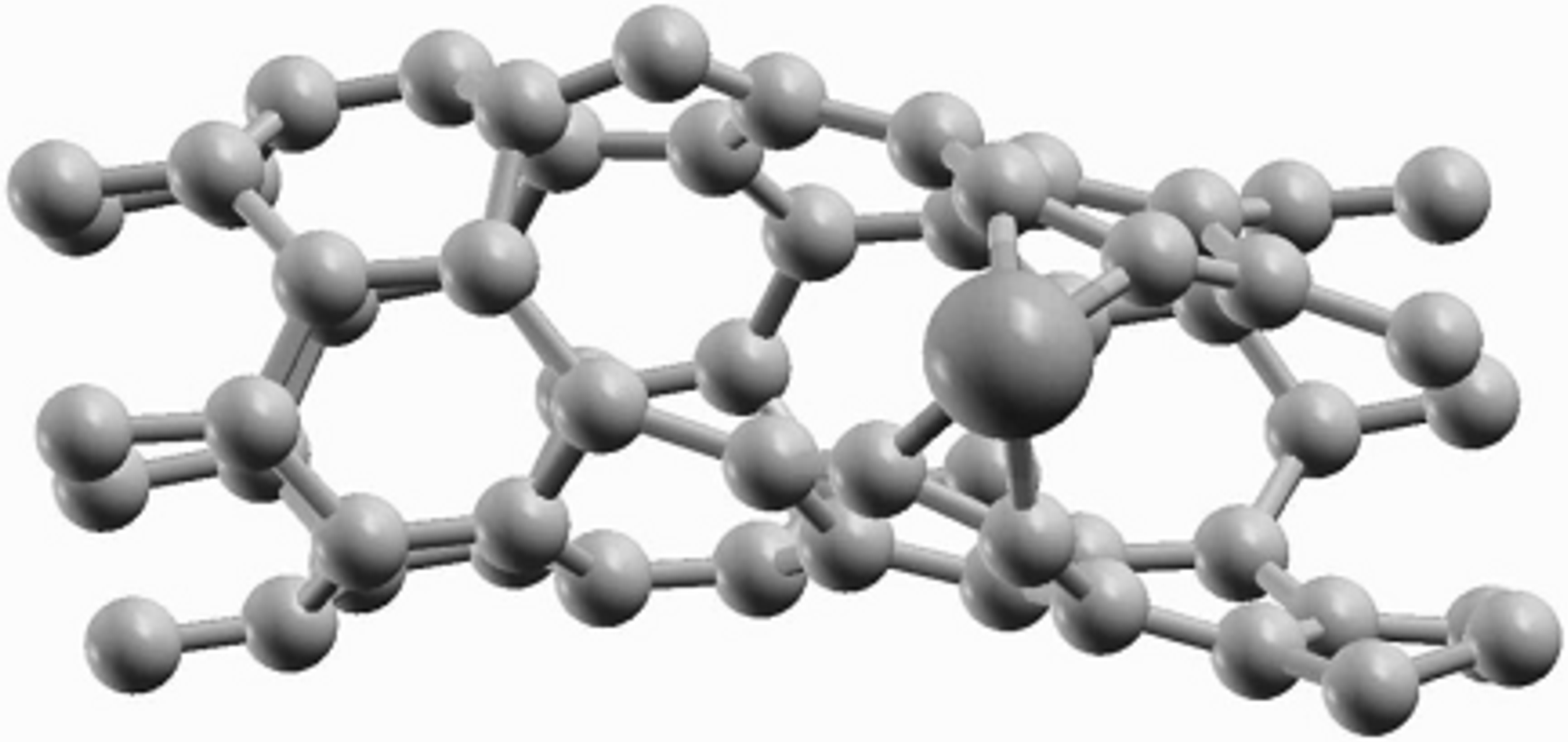} \\
(b) 
\includegraphics[width=0.8\hsize]{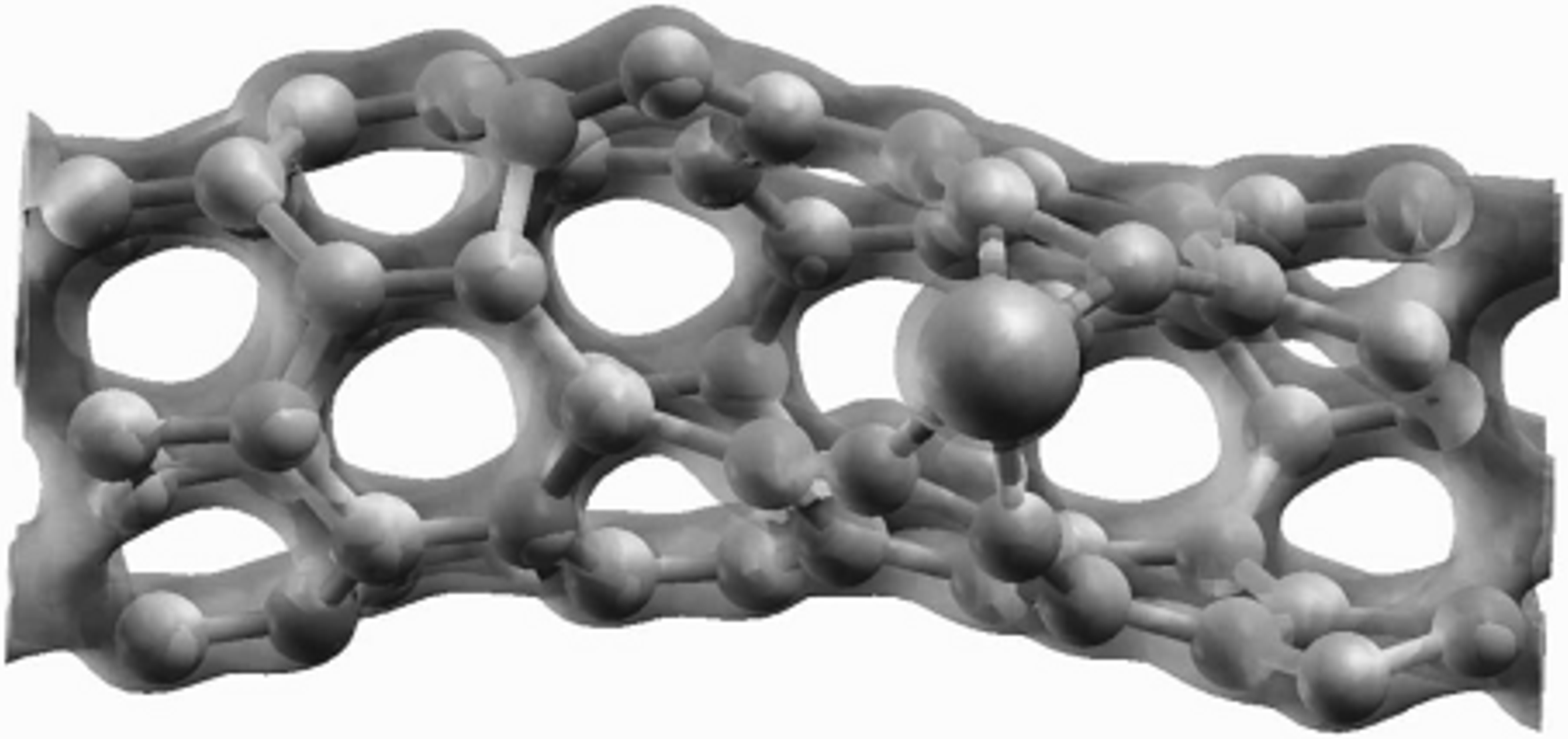}
\end{center}
\caption{(a) Adsorption of a tungsten atom on a twisted (5,0) nanotube. 
When the W atom (a big sphere) is attached on a tube wall, 
a big displacement of carbon atoms 
happens with opening of the tube wall. 
In (B), iso-surface of the charge distribution is overwritten. }
\label{f2}
\end{figure}

Enhancement of reactivity of the tube wall is much clearly seen, 
if we consider a twisted nanotube. 
In Fig.~\ref{f2}, we show an optimized structure obtained 
by adsorption of a tungsten atom on a twisted (5,0) nanotube. 
A big motion of carbon atoms 
happens and finally we have an opening of a breakage in the tube wall. 
In this process, the valence charge density is completely redistributed. 
As clearly seen in Fig.~\ref{f2}, several CC bonds are broken. 

In Table~\ref{t2}, 
we summarize adsorption energy of a metal on various nanotubes 
considered in this study. 
If we look at the adsorption energy on the twisted nanotube, 
a clear difference is seen between transition metal elements and 
noble metal elements. 
This is caused by existence or non-existence of a metal complex 
in the optimized structure. 
For the case of W, Ta, Nb, and Mo, formation of this complex is obtained. 
But, Pt and Au only form bondings with two carbon atoms at the adsorbed site. 
There is no bond-breaking in the carbon network, although 
the twisted (5,0) tube is in a rather unstable structure. 
This tendency in chemical trend 
holds also for the nanotube with SWD and 
roughly for the pristine nanotube. 

For analysis of the chemical nature of adsorption, 
the L\"{o}wdin charge is obtained in our simulation. 
For transition-metal elements preferring stronger chemisorption 
than noble metals, compared to an isolated atom, 
an atom adsorbed on a tube loses electrons. 
In this change, charge from $s$ electrons decreases significantly, 
while $p$ and $d$ electrons increase. 
However, for the case of Pt adsorption, 
the total charge rather increases on the Pt atom. 
(See Table~\ref{t3}.) 
The amount of decreased $s$ electrons is not so large, 
while slight increase in $p$ and $d$ electrons are found. 
Adsorption sites for Pt are 
the A site, a three-fold coordinated site, 
and the Z site (above a zigzag CC bond), for 
the pristine, defected with SWD, and twisted (5,0) tubes, respectively. 
We find less site-dependency in the charge transfer. 
Thus, the chemical trend on the charge redistribution can be attributed 
to the formation of metal-carbon complexes. 
Compared to noble metal atoms, which has an almost closed $d$ shell, 
transition metal elements tend to form 
the metal-carbon complex, and consequent strong chemical reactivity 
with deformed carbon nanotubes. 

If we refer to the strain release mechanism 
known in experiments as introduced in the introduction, 
we may not have an easy way to obtain 
a similar twisted tube in real experiments. 
However, when the tube is in a strained condition, 
formation of SWD is expected, for which 
we have shown enhanced reactivity, too. 
Even if the local strain is finally released by the formation of SWD, 
we can consider an intermediate meta-stable structure instantaneously. 
Since defect formation due to the local stress could be a rare event 
in a sub-microsecond phenomenon, a structure like the twisted nanotube 
prepared in our simulation might be realized in nature. 
Thus, we conclude that similar metal-carbon reactions should 
happen in deformed nanotubes rather in general. 

\section{Conclusions and discussion}

In this study, 
effects of single-metal adsorption on deformed nanotubes 
are investigated. 
Reactivity of the tube wall is shown to be high for 
tungsten, tantalum, and niobium. 
These late transition-metal elements with 5d or 4d electrons 
have rather large electron affinity. 
But, when platinum or gold, which have comparable large affinity, 
is adsorbed, similar reaction was not found, 
even if deformed tube walls were considered. 
The origin of the chemical trend is 
thus formation of metal-carbon complexes 
found for transition-metal elements. 

Since we have shown clear difference 
between these transition-metal elements and noble metals, 
global chemical trends on the reactivity is concluded. 
If one can adsorb W, Ta, or Nb atoms on a deformed tube, 
these particles would go to the defective sites, 
where metal-carbon complex are easily created, as shown in our simulations. 

We have considered a twisted nanotube. 
The reaction with these transition-metal atoms is much enhanced, 
when the deformed tube is used. 
The opening of a big breakage is observed in simulation results. 
Again, creation of local metal-carbon complex is a trigger for 
this remarkable reaction. 
This enhancement by mechanical modification is expected, 
even when one uses thicker nanotubes. 

Owing to these interesting phenomena found in our simulations, 
we can hope the enhanced reaction to realize next 
applications. 
\begin{description}
\item[Identifier for SWD] 
Positions of SWD can be detected by 
stable adsorbants, {\it i.e.} W, Nb, or Ta atoms, 
at the defect sites. 
Direct observation by electron microscope, or STM, 
as well as measurements using local probes, would be utilized. 
\item[Nanotube cutter]
In a solution, we can prepare W, Nb, or Ta ions easily by utilizing 
{\it e.g.} the Suzuki-Miyaura cross coupling catalysts.\cite{Suzuki-Miyaura} 
When a torsion is applied for nanotube in the solution, 
the tube wall can be broken. Thus, the catalytic molecule 
can be used as a nanotube cutter. 
The tube may be in a bundle of tubes. 
\item[Creation of nano-graphene]
After breakage of the tube wall occurs, 
the solution can be changed into a reductive environment. 
The complex structure in the metal adsorbed nanotube may be reduced. 
In this process, edges of the breakage in 
the carbon $\pi$ network might be hydrogenated. 
At the same time, the adsorbed metal atom may be reduced 
and it would be released from the carbon structure. 
Then, we will have a product of an opened nano-carbon structure 
from the closed nanotube. 
The broken wall of a tube will produce graphene flakes in pieces. 
A possible advantage of the method might be mass production of 
selected nano-graphene structures,\cite{Graphene,Fujita} 
if we could obtain nanotubes 
in a single chirality. 
\item[Test for mechanical strength of a tube]
The above redox process could be an experimental method to test 
mechanical stability of the nanotube, by finding 
products, and by measure the rate of reactions. 
\end{description}

\acknowledgment

This work is partly supported by a Grant-in-Aid for Scientific Research: 
KAKEN-HI (Grant No. 19054013 and No. 19310094), the Global COE Program 
and the Grand Challenges to Next-Generation Integrated Nanoscience. 

\newpage

\newpage

\newpage

\ \\

\begin{table}
\caption{Inter-atomic distances in metal adsorbed nanotube with 
Stone-Wales transformations. CC1 represents a carbon-carbon 
distance at a bond forming a 5-membered ring and a 7-membered ring, 
and CC2 represents that of a bond forming 
a 6-membered ring and the 7-membered ring. 
The adsorbed atom locates around this 7-membered ring. 
The mean distance from an adsorbed metal atom to carbon atoms 
is denoted by CM. }
\label{t1}
\begin{tabular}{|c|c|c|c|}
\hline
Metal & CC1 [\AA] & CC2 [\AA] & CM [\AA] \\
\hline
None & 1.465 & 1.458 & -     \\
Nb   & 2.638 & 3.013 & 2.171 \\
Mo   & 1.564 & 1.603 & 2.211 \\
Ta   & 2.663 & 3.091 & 2.148 \\
W    & 2.619 & 3.020 & 2.108 \\
Pt   & 1.487 & 1.520 & 2.159 \\
Au   & 1.505 & 1.453 & 2.139 \\
\hline
\end{tabular}
\end{table}

\begin{table}
\caption{Adsorption energy ([eV]) of a metal on various nanotubes. 
Pr denotes the pristine (5,0) nanotube, SW represents a tube with 
the Stone-Wales transformation, and Tw is a twisted (5,0) nanotube. 
See the text.}
\label{t2}
\begin{tabular}{|c|c|c|c|}
\hline
Metal & Pr &  SW &  Tw \\
\hline
Nb   & 3.48 & 6.90 & 17.25 \\
Mo   & 4.42 & 4.41 & 15.36 \\
Ta   & 3.41 & 8.52 & 16.17 \\
W    & 4.45 & 7.31 & 16.27 \\
Pt   & 2.65 & 2.23 & 2.32 \\
Au   & 0.94 & 0.68 & 1.61 \\
\hline
\end{tabular}
\end{table}

\newpage

\begin{table}
\caption{
Change in 
the L\"{o}wdin charges given in the pseudo-potential calculations 
for some metal atoms between isolated and adsorbed conditions. 
Difference in the value from charges for isolated atoms are given. 
Charges assigned for the valence orbitals ($s$, $p$, or $d$) 
in addition to the total L\"{o}wdin charge are shown. 
Pr, Sw, and Tw represent adsorption on the pristine (5,0) tube, 
the (5,0) tube with SWD, and the twisted (5,0) tube, respectively. 
}
\label{t3}
\begin{tabular}{|cc|c|c|c|c|}
\hline
Metal & Tube & Total &  s &  p & d \\
\hline
Nb 
  & Pr & -0.40 & -1.10 & +0.27 & +0.42 \\
  & Sw & -0.45 & -1.07 & +0.45 & +0.15 \\
  & Tw & -0.36 & -1.09 & +0.35 & +0.37 \\
  \hline
Mo 
  & Pr &  -0.24 & -0.85 & +0.28 & +0.34 \\
  & Sw &  -0.29 & -0.88 & +0.31 & +0.29 \\
  & Tw &  -0.22 & -0.76 & +0.49 & +0.05 \\
  \hline
W 
  & Pr & -0.34 & -1.39 & +0.50 & +0.55 \\
  & Sw & -0.36 & -1.34 & +0.61 & +0.36 \\
  & Tw & -0.26 & -1.36 & +0.54 & +0.57 \\
  \hline
Pt 
  & Pr & +0.05 & -0.28 & +0.21 & +0.13 \\
  & Sw & +0.01 & -0.37 & +0.29 & +0.08 \\
  & Tw & +0.08 & -0.32 & +0.28 & +0.13 \\
  \hline
\end{tabular}
\end{table}


\begin{thebibliography}{9}
\bibitem{Iijima} S. Iijima: Nature {\bf 354} (1991) 56. 
\bibitem{Thess} A. Thess, R. Lee, P. Nikolaev, H. Dai, P. Petit, J. Robert, C. Xu, Y.H. Lee, S.G. Kim, A.G. Rinzler, D.T. Colbert, G.E. Scuseria, D. Tom\'{a}nek, J.E. Fischer, R.E. Smalley: Science {\bf 273} (1996) 483.
\bibitem{Tans} S.J. Tang, A.R.M. Verschueren, C. Dekker: Nature, {\bf 393} (1998) 49.
\bibitem{Martel} R. Martel, T. Schmidt, H.R. Shea, T. Hertel, and Ph. Avouris: Appl. Phys. Lett. {\bf 73} (1998) 2447.
\bibitem{Zhang} Y. Zhang, N.W. Franklin, R.J. Chen, and H. Dai: Chem. Phys. Lett. {\bf 331} (2000) 35. 
\bibitem{Jin-Phillipp} N.Y. Jin-Phillipp and M. R\"{u}hle: Phys. Rev. B {\bf 70} (2004) 245421.
\bibitem{Yang2} S.H. Yang, W.H. Shin, J.W. Lee, H.S. Kim, J.K. Kang, and Y.K. Kim: Appl. Phys. Lett. {\bf 90} (2007) 013103. 
\bibitem{Durgun} E. Durgun, S. Dag. V.M.K. Bagci, O. G\"{u}lseren, T. Yildirim, and S. Ciraci: Phys. Rev. B {\bf 67} (2003) 201401(R). 
\bibitem{Yang1} S.H. Yang, W.H. Shin, J.W. Lee, S.Y. Kim, S.I. Woo, and J.K. Kang: J. Phys. Chem. B {\bf 110} (2006) 13941. 
\bibitem{Zhuang} H.L. Zhuang, G.P. Zheng, and A.K. Soh: Comput. Mat. Sci. {\bf 43} (2008) 823. 
\bibitem{Duong} H.M. Duong, D.V. Papavassiliou, K.J. Mullen, B.L. Wardle, and S. Maruyama: J. Phys. Chem. C {\bf 112} (2008) 19860. 
\bibitem{Inoue} S. Inoue and Y. Matsumura: Chem. Phys. Lett. {\bf 464} (2008) 160. 
\bibitem{David} M. David, T. Kishi, M. Kisaku, H. Nakanishi, and H. Kasai: Surf. Sci. {\bf 601} (2007) 4366. 
\bibitem{Hohenberg} P. Hohenberg and W. Kohn: Phys. Rev. {\bf 136} (1964) B864.
\bibitem{Kohn}W. Kohn and L.J. Sham: Phys. Rev. {\bf 140} (1965) A1133.
\bibitem{Giannozzi} 
P. Giannozzi, S. Baroni, N. Bonini, M. Calandra, R. Car, 
C. Cavazzoni, D. Ceresoli, G. L Chiarotti, M. Cococcioni, 
I. Dabo, A. Dal Corso, S. de Gironcoli, S. Fabris, 
G. Fratesi, R. Gebauer, U. Gerstmann, C. Gougoussis, 
A. Kokalj, M. Lazzeri, L. Martin-Samos, N. Marzari, 
F. Mauri, R. Mazzarello, S. Paolini, A. Pasquarello, 
L. Paulatto, C. Sbraccia, S. Scandolo, G. Sclauzero, 
A. P Seitsonen, A. Smogunov, P. Umari, and R. M Wentzcovitch: 
J. Phys.: Condens. Matter {\bf 21} (2009) 395502.
\bibitem{Perdew2} 
J.P. Perdew, K. Burke, and M. Ernzerhof: Phys. Rev. Lett. {\bf 77} (1996) 3865. 
\bibitem{Perdew3} 
J.P. Perdew, K. Burke, and M. Ernzerhof: Phys. Rev. Lett. {\bf 78} (1997) 1396.
\bibitem{Vanderbilt} D. Vanderbilt: Phys. Rev. B {\bf 41} (1990) 7892. 
\bibitem{Ebbessen} T.W. Ebbsesen and T. Takada: Carbon {\bf 33} (1995) 973.
\bibitem{Charlier-rev} J.-C. Charlier: Acc. Chem. Res. {\bf 35} (2002) 1063.
\bibitem{Jiang} H. Jiang, X.-Q. Feng, Y. Huang, K.C. Hwang, and P.D. Wu: Comput. Methods Appl. Mech. Eng. {\bf 193} (2004) 3419.
\bibitem{Stone} A.J. Stone and D.J. Wales: Chem. Phys. Lett. {\bf 128} (1986) 501. 
\bibitem{Suenaga} K. Suenaga, H. Wakabayashi, M. Koshino, Y. Sato, K. Urita, and S. Iijima: Nature Nanotechnology {\bf 2} (2007) 358.
\bibitem{Nardelli1} M.B. Nardelli, B.I. Yakobson, and J. Bernholc: Phys. Rev. Lett. {\bf 81} (1998) 4656. 
\bibitem{Nardelli2} M.B. Nardelli, B.I. Yakobson, and J. Bernholc: Phys. Rev. B {\bf 57} (1998) R4277. 
\bibitem{Terrones} H. Terrones, M. Terrones, E. Hern\'{a}ndez, N. Grobert, J.-C. Charlier, and P.M. Ajayan: Phys. Rev. Lett. {\bf 84} (2000) 1716.
\bibitem{Chandra} N. Chandra, S. Namilae, and C. Shet: Phys. Rev. B {\bf 69} (2004) 094101. 
\bibitem{Wako} K. Wako, T. Oda, M. Tachibana, and K. Kojima: Jpn. J. Appl. Phys. {\bf 47} (2008) 6601.
\bibitem{Pozrikidis} C. Pozrikidis: Arch. Appl. Mech. {\bf 79} (2009) 113. 
\bibitem{Amelinckx} S. Amelinckx, X.B. Zhang, D. Bernaerts, X.F. Zhang, V. Ivanov, and J.B. Nagy: Science {\bf 265} (1994) 635.
\bibitem{Charlier} J.-C. Charlier, T.W. Ebbesen, and Ph. Lambin: Phys. Rev. B {\bf 53} (1996) 11108.
\bibitem{Carroll} D.L. Carroll, P. Redlich, P.M. Ajayan, J.C. Charlier, X. Blase, A. De Vita, and R. Car: Phys. Rev. Lett. {\bf 78} (1997) 2811.
\bibitem{Kim} P. Kim, T.W. Odom, J.-L. Huang, and C.M. Lieber: Phys. Rev. Lett. {\bf 82} (1999) 1225. 
\bibitem{Meunier} V. Meunier and Ph. Lambin: Carbon {\bf 38} (2000) 1729. 
\bibitem{Tekleab} D. Tekleab, D.L. Carroll, G.G. Samsonidze, and B.I. Yakobson: Phys. Rev. B {\bf 64} (2001) 035419. 
\bibitem{Yang-Shin-Kang} S.H. Yang, W.H. Shin, and J.K. Kang: J. Chem. Phys. {\bf 125} (2006) 084705.
\bibitem{Horner} D.A. Horner, P.C. Redfern, M. Sternberg, P. Zapol, and L.A. Curtiss: Chem. Phys. Lett. {\bf 450} (2007) 71.
\bibitem{Yakobson} B.I. Yakobson, C.J. Brabec, and J. Bernholc: Phys. Rev. Lett. {\bf 76} (1996) 2511. 
\bibitem{Wong} E.W. Wong, P.E. Sheehan, and C.M. Lieber: Science {\bf 277} (1997) 1971. 
\bibitem{Seo} K. Seo, K.A. Park, C. Kim, S. Han, B. Kim, and Y.H. Lee: J. Am. Chem. Soc. {\bf 127} (2005) 15724.
\bibitem{Miyake} T. Miyake and S. Saito: Phys. Rev. B {\bf 68} (2003) 155424.
\bibitem{Conversano} R. Conversano, F. Cleri, G. D'Agostino, V. Rosato, and M. Volpe, in Nanotubes and Related Materials: MRS Symposium Proceedings No. 633 (Materials Research Society, Pittsburgh, 2001), p. F14.8.1.
\bibitem{Zhou} L.G. Zhou and S.-Q. Shi: Appl. Phys. Lett. {\bf 83} (2003) 1222.
\bibitem{Picozzi} S. Picozzi, S. Santucci, L. Lozzi, L. Valentini, and B. Delley: J. Chem. Phys. {\bf 120} (2004) 7147.
\bibitem{Li} L. Li, S. Reich, and J. Robertson: Phys. Rev. B {\bf 72} (2005) 184109.
\bibitem{Suzuki-Miyaura} N. Miyaura, and A. Suzuki: J. Chem. Soc., Chem. Commun. (1979) 866.
\bibitem{Graphene} K.S. Novoselov, A.K. Geim: Nature, {\bf 438} (2005) 197.
\bibitem{Fujita} M. Fujita, K. Wakabayashi, K. Nakada, and K. Kusakabe: J. Phys. Soc. Jpn. {\bf 65} (1996) 1920.
\end{thebibliography}
\end{document}